\documentclass[11pt,a4paper]{article}
\usepackage{jheppub}
\relax

\def\bes{\begin{eqnarray}}
 \def\ees{\end{eqnarray}}
\def\be{\begin{equation}}
\def\ee{\end{equation}}
\def\bs{\begin{subequations}}
\def\es{\end{subequations}}
\newcommand{\een}{\end{subequations}}
\newcommand{\ben}{\begin{subequations}}
\newcommand{\beq}{\begin{eqalignno}}
\newcommand{\eeq}{\end{eqalignno}}

\def\calP{{{\cal{P}}}}
\def\Et{{\tilde{E}}}

\def\yt{{\tilde{y}}}
\def\tit{{\tilde{t}}}
 \def\ex{\epsilon}
 \def\lx{\lambda}
\def\et{{\tilde{\ex}}}

\def\rt{{\tilde{r}}}

\def\phit{{\tilde{\phi}}}
\def\ct{{\tilde{c}}}

\title{Shock waves as branes with throats}
\author[1]{J. Rizos}
\author[2]{and N. Tetradis}
\affiliation[1]{Department of Physics,\\ University of Ioannina, Ioannina 45110, Greece}
\affiliation[2]{Department of Physics,\\ University of Athens, Zographou 15784, Greece}

\emailAdd{irizos@uoi.gr}
\emailAdd{ntetrad@phys.uoa.gr}

\abstract{
We discuss the properties of a class of exact dynamical solutions of the
DBI action in various dimensions. They can be interpreted as shock waves of the 
nonlinear theory. They can also describe two parallel branes connected
by a throat within a Minkowski bulk.  We analyze issues related to the conservation 
of energy in these systems. We also discuss possible applications in 
particle physics and cosmology. 
}


\begin{document}
\maketitle

\section{Introduction}\label{intro}

Scalar field theories with derivative interactions have been discussed extensively during the last years in the context of 
particle physics and cosmology. Depending on the form of the Lagrangian, the underlying symmetries 
and the particular application, they go under a variety of names, such as: $k$-essence \cite{kessence}, Dirac-Born-Infeld (DBI) 
inflation \cite{dbin},
the Dvali-Gabadadze-Porrati (DGP) model \cite{dgp} in the decoupling limit and the Galileon \cite{galileon}, 
scalar-tensor models with kinetic gravity braiding \cite{braiding}, etc. 
A common property of these theories is that 
the equation of motion does not contain field derivatives higher than the second. As a result, ghost fields do not appear
in the spectrum.  
The most general scalar-tensor
theory with this property was constructed a long time ago \cite{horndeski}, and rediscovered recenty in the context of
the generalized Galileon (see \cite{genegal} and references therein).

An alternative perspective can be obtained by constructing a geometric 
picture of some of these theories. 
As shown in \cite{dbigal}, the Galileon theory can be reproduced by considering 
the position modulus of a probe brane within a (4+1)-dimensional bulk. This geometric approach demonstrates how derivatives 
higher than the second in the effective theory can be eliminated by employing the Lovelock invariants.
The presence of derivative interactions, besides complicating the discussion of quantum corrections, generates 
nontrivial features already at the classical level \cite{quantum}. 

The purpose of the present work is to discuss the properties of a class of solutions in the context of the
$(d+1)$-dimensional DBI theory. Even though their form in 3+1 dimensions has already been presented in \cite{dynamical}, 
their features have not been analyzed. We show that the generalized solution involves a singularity 
that can be viewed as a propagating shock wave in 
$d+1$ dimensions. This interpretation has been put forward by Heisenberg for an effective $(1+1)$-dimensional
theory in order to describe meson production through nucleon scattering \cite{heisenberg}.
Energy is transferred from the location of the shock front, where the energy density diverges, 
to the region behind it. During this process the singularity acts as an energy source. 
One very interesting
feature of the solution is that the shock wave cannot evolve below a minimal radius not related to the
fundamental scale of the theory.

The DBI theory can be viewed as the effective description of a $(d+1)$-dimensional brane embedded in 
a Minkowski bulk space with one additional spatial dimension.
In this context, our solution can be used in order to construct a
time-dependent generalization of the static catenoidal configuration of Callan, Maldacena \cite{callan} and Gibbons \cite{gibbons}.   
In this work we consider only the case with vanishing gauge fields.
The configuration is obtained by joining two solutions with opposite field derivatives at the point where they
display square-root singularities. The result is a smooth surface that looks like a throat or wormhole 
connecting two asymptotically parallel branes. The radius of the throat is time dependent and has a minimal value that
is independent of the fundamental length scale. The brane picture provides a link between the energy 
generated by the singularity and the worldvolume of the branes that is eliminated through the expansion of the throat.

We should note that we do not identify the branes with the D-branes of string theory. We treat them simply as 
geometrical defects in a higher-dimensional space in the probe approximation. In particular, we 
neglect any long-range interaction between the branes as well as string processes when they approach each other. 
We use the brane picture mainly in order to resolve the question of energy conservation in the 
shock-wave picture. 

In the following section we present the solutions and study their energetics. In section \ref{sf} we discus their interpretation 
as shock waves and in section \ref{bt} as branes joined by throats. In section \ref{gaugee} we analyze the brane picture
through a different choice of worldvolume coordinates that eliminate the singularity from the location of the throat. 
In section \ref{concl} we give our conclusions and discuss possible extensions of the results.

\section{The solutions}\label{solut}

We consider a theory in $d+1$ dimensions described by a Lagrangian density of the
Dirac-Born-Infeld (DBI) type:
\be
{\cal L}=-\frac{1}{\lx}\sqrt{1-\lx \left(\partial_\mu\phi \right)^2}+\frac{1}{\lx}. 
\label{lagrangian} \ee
For most of the paper we shall assume that $\lx>0$, with the fundamental length scale of the theory being $\sim \lx^{1/d}$.
However, we shall also allow the possibility $\lx<0$, to which we refer as the ``wrong"-sign DBI theory. 
We have included a constant term $1/\lx$ in the action so that the energy density vanishes
for configurations with $\phi=0$. As a result, the theory does not contain a cosmological constant.
The equation of motion of the field $\phi$ is
\be
\partial^\mu\left[ \frac{  \partial_\mu \phi}{\sqrt{1-\lx\left(\partial_\nu\phi \right)^2}}\right] =0.
\label{eom} \ee
Its form makes apparent the existence of a Noether current, arising from the invariance of the Lagrangian under the
shift $\phi\to\phi+c$. 
The energy-momentum tensor is
\be
T^{\mu\nu}=-\frac{  \partial^\mu \phi  \partial^\nu \phi}{\sqrt{1-\lx\left(\partial_\rho\phi \right)^2}}
-\eta^{\mu\nu} \frac{1}{\lx}\left(\sqrt{1-\lx\left(\partial_\rho \phi \right)^2}-1\right).
\label{tmn} \ee

We consider spherically symmetric field configurations. 
When expressed in spherical coordinates, eq. (\ref{eom}) takes the form
\be
\partial_t\left[ \frac{\phi_t }{\sqrt{1- \lx \phi_t^2+\lx\phi_r^2 }} \right]
-\frac{1}{r^{(d-1)}}\partial_r\left[r^{(d-1)}\frac{\phi_r }{\sqrt{1- \lx \phi_t^2+\lx\phi_r^2 }}\right]
=0,
\label{eomsph} \ee
where subscripts denote partial derivatives.
We assume
$1- \lx \phi_t^2+\lx\phi_r^2 \geq 0$, 
an obvious constraint imposed by the form of the Lagrangian density (\ref{lagrangian}).
From the energy-momentum tensor we deduce the energy density $\rho$, 
radial energy flux $j_r$, radial pressure $p_r$ and angular pressure $p_\theta$.
They are given by 
\begin{eqnarray}
\rho&=& \frac{1+\lx\phi_r^2 }{\lx\sqrt{1- \lx \phi_t^2+\lx\phi_r^2 }}-\frac{1}{\lx}
\label{density} \\
j_r&=&-\frac{\phi_r\phi_t }{\sqrt{1- \lx \phi_t^2+\lx\phi_r^2 }}
\label{flux} \\
p_r&=& -\frac{1-\lx\phi_t^2 }{\lx\sqrt{1- \lx \phi_t^2+\lx\phi_r^2 }}+\frac{1}{\lx}
\label{pr}\\
p_\theta&=&\frac{1}{\lx}\left(1-\sqrt{1- \lx \phi_t^2+\lx\phi_r^2 } \right),
\label{pth}
\end{eqnarray} 
respectively.

We are interested in exact analytical solutions of eq. (\ref{eomsph}). 
 For $\phi=\phi(r)$, eq. (\ref{eomsph}) gives
$\phi_r/\sqrt{1+\lx\phi^2_r}=\pm c/r^{d-1}$,
where we have assumed that the constant of integration $c$ is positive.
This relation can be written as 
\be
\phi_r=\pm \frac{c}{\sqrt{r^{2(d-1)}-\lx c^2}}.
\label{stat2} \ee
For the ``wrong"-sign DBI theory with $\lx<0$, the solutions extend down to $r=0$ for both signs.  
For $d=3$, 
they were interpreted in \cite{dvali1}  as field configurations induced by a $\delta$-function source resulting from the large concentration of
energy within a small region of space around $r=0$. 
An alternative interpretation was given in \cite{gibbons} in the context of the Born-Infeld theory. In this case the field corresponds
to the nonlinear electrostatic potential induced by a point charge. The corresponding 
configuration was termed BIon. Multi-Bion solutions were also constructed in the form of a periodic lattice.  

We are interested mainly in the solutions for 
$\lx>0$, which display 
square-root singularities at $r_{th}=(\lx c^2)^{1/(2d-2)}$. 
Folllowing \cite{gibbons} we can join smoothly the two branches of 
eq. (\ref{stat2}) with opposite signs, in order to create 
a continuous double-valued function of $r$ that extends from infinite $r$ to $r_{th}$ and back out to infinity. This catenoidal 
solution describes a pair of static branes embedded in $(d+1)$-dimensional Euclidean space, which are connected by
a throat. 

In the following we study similar configurations that evolve with time.
Exact time-dependent solutions of the (3+1)-dimensional theory, 
of the form $\phi=\phi(z)$, with $z=r^2-t^2$, were presented in \cite{dynamical}. They can 
be generalized to $d+1$ dimensions. 
Substituting the ansatz $\phi=\phi(z)$ in eq. (\ref{eomsph}) gives 
\be
2 z \frac{d\phi_z(z)}{dz} +4 d \lx \, z\, \phi_z^3+ (d+1) \phi_z=0,
\label{hzeq} \ee
where the index denotes differentiation with respect to $z$. 
The solutions of this equation are 
 \be
\phi_z(z)=\pm\frac{\ct}{\sqrt{z^{d+1}-4\lx \ct^2 z}},
\label{hz2}
\ee
where we have defined the integration constant $\ct>0$ in a way that the solutions remain real for $r\to\infty$. 
For both signs of $\lx$ the solutions display square-root
singularities at the value $z_{th}$ that satisfies $z^d_{th}=4\lx \ct^2$ and at $z=0$.

The solutions (\ref{hz2}) can be interpreted as the time-dependent generalization of the static 
solutions (\ref{stat2}).
The two sets of solutions can also be related through analytic continuation. 
By continuing $t$ to imaginary values ($t=i \tau$) and performing the 
substitutions $z=r^2+\tau^2\to r^2$, $d\to d-1$, $\ct\to c/2$, the solutions (\ref{hz2}) reproduce (\ref{stat2}).
However, it must be emphasized that physically the time-dependent solutions are not a mere boosted version of the
static ones. For example, it can be seen easily that the integration constants $c$ and $\ct$ have different dimensions in 
(\ref{stat2}) and (\ref{hz2}).

A second class of solutions can be obtained if the configuration is confined within the region
$r^2\leq t^2$. By defining $w=t^2-r^2$, we obtain solutions of the form
 \be
\phi_w(w)=\pm\frac{\ct}{\sqrt{\pm w^{d+1}+4\lx \ct^2 w}},
\label{hz2b}
\ee
where we assume again that $\ct>0$.
Depending on the value of $d$, the solutions with one of the two signs under the square root are a 
rewriting of eq. (\ref{hz2}). The other sign results in a new set of solutions. The shock-wave configuration discussed
in \cite{heisenberg} belongs to this class. However, we would like to maintain the connection with the picture of 
brane dynamics. For this reason we focus on the solutions (\ref{hz2}), for which the field remains nonzero
at large distances and vanishes only asymptotically.

The energy density and flux, and the radial and angular pressures are given by eqs. (\ref{density})-(\ref{pth}).
For the solution (\ref{hz2}) they become
\begin{eqnarray}
\rho&=&\frac{1}{\lx}\left( \frac{ u^{d+1}+\tit^2}{u^{(d+2)/2}\sqrt{u^d-1}}-1 \right)
\label{rhodd} \\
j_r&=&\frac{1}{\lx}\frac{\tit \sqrt{u+\tit^2}}{u^{(d+2)/2}\sqrt{u^d-1}} 
\label{jjdd}  \\
p_r&=&\frac{1}{\lx}\left(1- \frac{ u^{d+1}-u-\tit^2}{u^{(d+2)/2}\sqrt{u^d-1}} \right)
\label{ppdd} \\
p_\theta&=&\frac{1}{\lx}\left( 1- \frac{ u^{d/2}}{\sqrt{u^d-1}} \right),
\label{pthdd} \end{eqnarray}
where $z=(4\lx \ct^2)^{1/d}u$, $t^2=(4\lx \ct^2)^{1/d}\tit^2$, $r^2=(4\lx \ct^2)^{1/d}\rt^2$, $u=\rt^2-\tit^2$.
The energy density and flux satisfy the continuity equation 
$\partial\rho/\partial t+(1/r^{d-1})\partial(r^{d-1}j_r)/\partial r=0$, resulting from the conservation of the
energy-momentum tensor (\ref{tmn}). 
All the above quantities diverge at the location of the singularity $r_{th}(t)=\sqrt{(4\lx \ct^2)^{1/d}+t^2}$.

At a given time $t$, the total energy of the configuration contained within the volume outside the singularity is
\begin{eqnarray}
E_{th}=\Omega_{d-1} \int_{r_{th}(t)}^\infty  r^{d-1} \rho(t,r) dr &=& 
\frac{ \ct}{\sqrt{\lx}}\Omega_{d-1} \int_{1}^\infty
\left( \frac{ u^{d+1}+\tit^2}{u^{(d+2)/2}\sqrt{u^d-1}}-1 \right) \left( u+\tit^2 \right)^{(d-2)/2} \, du
\nonumber \\
&=&\frac{\ct}{ \sqrt{\lx}}  \Omega_{d-1} \frac{2}{d}  \left(1+\tit^2 \right)^{d/2},
\label{ethroatdd} \end{eqnarray}
where $\Omega_{d-1}=2 \pi^{d/2}/\Gamma(d/2)$ is the area of the $(d-1)$-sphere.
The energy is finite because the square-root singularity is integrable. On the other hand, $E_{th}$ is time dependent,
which seems to violate the notion of conservation of energy.
As we have mentioned, the energy density and flux satisfy the continuity equation that results from the conservation of the
energy-momentum tensor (\ref{tmn}). Also, the energy flux vanishes at radial infinity. 
It is clear then that the time dependence of $E_{th}$ is related to the presence of a singularity in the solution, 
at a point where the energy density and flux diverge. 

In the following sections we shall analyze this issue from two different points of view. 
Two observations are useful:
\begin{enumerate}
\item
In principle it could be possible to account for the time dependence of $E_{th}$ by
allowing the field to have a nontrivial profile in the region $r<r_{th}$. For the ansatz $\phi=\phi(w)$, with $w=t^2-r^2\geq 0 $,
the relevant solutions are given by eq. (\ref{hz2b}). However, they have a non-integrable singularity at
$r=t$, while they fail to remain real in the region $t<r<r_{th}(t)$. 
Another consistent solution has a constant $\phi(z)$. If one does not wish to introduce artificial $\delta$-function
singularities in the first derivatives of the full 
solution at $r=r_{th}$, the constant field value must be taken equal to the value resulting from the solution
(\ref{hz2}) at $r=r_{th}$. This construction does not introduce any new contributions to the energy. 
\item
We can recast eq (\ref{ethroatdd}) in the more intuitive form 
\be
E_{th}=\frac{1}{\lx}\frac{\Omega_{d-1}}{d}\, r^d_{th}(t).
\label{ethroatddd} \ee
The energy outside the singularity can be expressed as the product of the fundamental energy density $1/\lx$ and  
the volume $V_{d-1}$ of a $(d-1)$-sphere of radius $r_{th}=\sqrt{(4\lx c^2)^{1/d}+t^2}$.
\end{enumerate}

\section{Shock waves}\label{sf}

In \cite{heisenberg} the solution (\ref{hz2}) with $d=1$ was used by Heisenberg for the description of meson production in 
the collision of two nucleons. Because of the high boost factor the colliding nucleons were approximated as disk-like. For this reason the 
products of the collision were assumed to have initially a planar geometry. The DBI action was employed as a
phenomenological description of the underlying dynamics. 
The singularity in eq. (\ref{hz2}) was interpreted as a shock front, where the energy density diverges. It was observed that 
the solution can be expanded in terms of plane waves with an amplitude that increases $\sim t^{1/2}$, consistently with the
increase of the total energy $\sim t$ predicted by eq. (\ref{ethroatdd}) with $d=1$ at late times. With increasing time, energy is
transferred from a region of infinite density at the shock wave to its wake, where it is associated with the production of mesons.

Despite the intuitive interpretation of the solution, certain issues remain ambiguous:
\begin{enumerate}
\item
 The shock wave is not of the standard type, as the  energy density and pressure vanish
on one of its sides.
\item
As already remarked in \cite{heisenberg}, it is expected that the energy stored in the shock will be depleted eventually, when it 
is transferred entirely to the wake. The solution (\ref{hz2}) does not display this behavior and must be considered only as
an approximation. 
\end{enumerate}

In order to obtain some better understanding of how the local energy conservation fails to translate into the conservation
of a global quantity we look more carefully at the solution for $d=3$. 
If we adopt an intepretation similar to the planar case, the solution describes a spherical shock wave that moves in from
infinite to a minimal radius,  
while time flows from $-\infty$ to $0$, and then moves out again for positive $t$. 
The shock front is located at $z_{th}=(4\lx \ct^2)^{1/3}$, or at $r_{th}(t)=\sqrt{(4\lx \ct^2)^{1/3}+t^2}$. 
The field can be expressed as $\phi(z)=\phi_0+(\ct/\lx)^{1/3}{\cal P}^{-1}(s,0,1)$, where 
 ${\cal P}^{-1}$ is the inverse Weierstrass function and $s=(2^{1/3}z+(\lx\ct^2)^{1/3})/(z-(4\lx\ct^2)^{1/3})$.
If the field is assumed to vanish for $r\to \infty$, its value at the shock front is 
$\phi(t,r_{th}(t))=B_3 \left( {2\ct}/{ \lx}\right)^{1/3}$,
with $B_3=\sqrt{\pi}\,\Gamma\left({1}/{3}\right)/\left(6\,\Gamma(5/6)\right)\simeq 0.701$.
The partial derivatives $\partial \phi/\partial t$ and  $\partial \phi/\partial r$ diverge at $r=r_{th}(t)$.

The energy density $\rho$ and
radial energy flux $j_r$ for this solution are given by eqs. (\ref{rhodd}), (\ref{jjdd}) with $d=3$. 
The rescaled coordinates are 
$z=(4\lx \ct^2)^{1/3}u$, $t^2=(4\lx \ct^2)^{1/3}\tit^2$, $r^2=(4\lx \ct^2)^{1/3}\rt^2$, $u=\rt^2-\tit^2$.
At a given time $t$, the total energy of the configuration is 
\be
E_{th}=\int_{r_{th}(t)}^\infty 4\pi r^2 \rho(t,r) dr =\frac{8\pi \ct}{3 \sqrt{\lx}}\left(1+\tit^2 \right)^{3/2}.
\label{ethroat} \ee
The energy is finite, but time dependent. This result seems to contradict energy conservation.
However, it is easy to check that the energy density and flux satisfy the continuity equation 
$\partial\rho/\partial t+(1/r^2)\partial(r^2j_r)/\partial r=0$, resulting from the conservation of the
energy-momentum tensor (\ref{tmn}). Also, the energy flux vanishes at infinite radius. 
This means that the energy increase must be attributed to the presence of the singularity. However, no
energy seems to be contained within the volume enclosed by the wave, while there is no obvious 
indication for the presence of an energy source at the location of the singularity. 

In order to understand better this issue we consider the total energy between surfaces located at $r(t)=r_{th}(t)+\ex$
and radial infinity. We have
\begin{eqnarray}
\frac{dE_{th}(\ex)}{dt}&=&\int_{r_{th}(t)+\ex}^\infty \left. 4\pi r^2\frac{\partial \rho(t,r)}{\partial t} dr
- 4\pi r^2  \rho(t,r) \right|_{r=r_{th}(t)+\ex}\frac{dr_{th}(t)}{dt}
\nonumber \\
&=&\left. 4\pi r^2\left( j_r(t,r) -  \rho(t,r)\frac{dr_{th}(t)}{dt} \right)\right|_{r=r_{th}(t)+\ex},
\label{enep} \end{eqnarray}
where we have used the continuity equation. It is clear that the variation of the energy $E_{th}(\ex)$ depends
on the difference between the energy flux of the fluid associated with the energy-momentum tensor (\ref{tmn}) and
an effective flux generated by the motion of the surface at $r(t)=r_{th}(t)+\ex$. In the limit $\ex\to 0$ we expect these
two fluxes to cancel each other, as the configuration we are considering is limited within the
region $r\geq r_{th}(t)$. A straightforward evaluation gives
\begin{eqnarray}
\rho \frac{dr_{th}}{dt}&=&\frac{\tit\left(1+\tit^2\right)^{1/4}}{\sqrt{6}\lx}\frac{1}{\sqrt{\et}}
-\frac{\tit}{\lx\sqrt{1+\tit^2}}+{\cal O}\left( \sqrt{\et}\right)
\nonumber \\
j_r&=&\frac{\tit\left(1+\tit^2\right)^{1/4}}{\sqrt{6}\lx}\frac{1}{\sqrt{\et}}
+{\cal O}\left( \sqrt{\et}\right)
\label{lim} \end{eqnarray}
for  $\ex=(4\lx \ct^2)^{1/6}\,\et \to 0$.
The leading (divergent) contributions are identical in the above expressions. However, the subleading parts are not the same, with
their difference remaining nonzero in the limit $\et\to 0$.  
It can be verified easily that eq. (\ref{enep})
is consistent with eq. (\ref{ethroat}) in this limit.

The above result can also be stated as follows:
Surfaces enclosing a constant amount of energy between their location and radial infinity
coincide with the shock front at finite values of $\tau$. As a result, the shock front encloses different amounts of energy 
at various times.
On the other hand, the velocity of the front $dr_{th}/dt$ and that of the fluid at its location $j_r/\rho$ coincide, as
$\rho (dr_{th}/dt)/j_r=1+{\cal O}(\sqrt{\ex})$ for $\ex\to 0$. This means that the absorption or emission of energy cannot
be traced to the region with $r<r_{th}$.
It is the divergence of the energy density at the shock location that causes this peculiar behavior.
We also point out that allowing the surface to have a more general
time dependence, of the form $r(t)=r_{th}(t)+\ex f(t)$,  does not resolve the problem. The difference 
$j_r-\rho\, (dr/dt)$ is ${\cal O}(1)$ and does not involve $f(r)$.

It seems that the paradoxical features in the interpretation of the solution (\ref{hz2}) in the context of the
effective theory (\ref{lagrangian})
arise from the divergence 
of the energy density. The singularity must be viewed also as the source or sink of energy. A more satisfactory 
picture can be obtained if the solution is viewed as a description of a throat connecting two parallel branes, which is
the topic of the following section. 

A very interesting
feature of the solution we discussed is that the shock front cannot evolve below a minimal radius not related to the
fundamental scale of the theory. From this point of view, the configuration realizes the concept of 
classicalization \cite{dvali1,dvali2}. 
This term has been attributed to 
the appearance of a new scale in derivatively coupled theories, which persists in the classical limit $\hbar \to 0$ 
\cite{dvali1,dvali2}. 
In theories that display classicalization a highly energetic 
classical configuration is expected to scatter at distances comparable to the new scale. 
When this is  much larger than the fundamental length scale, short distances cannot be probed even with arbitrarily large
center-of-mass energies. 

The existence and detailed nature of the phenomenon are still uncertain \cite{dynamical,road}. It has been 
suggested that classicalization is connected to the presence of classical configurations in the spectrum (classicalons), 
which are sourced by the energy. They correspond to the static solutions (\ref{stat2}) that we discussed earlier, with
$d=3$ and $\lx <1$. 
The physics of classicalization was attributed to the creation of such
field configurations during high-energy scattering \cite{dvali1}. 
On the other hand, spherical collapsing configurations were studied numerically in \cite{dynamical} without observing the 
creation of classicalons or significant scattering. Despite the use of the term classicalization, it is possible that
quantum physics must play a crucial role in the realization of this idea \cite{road}.

The time-dependent solution (\ref{hz2}), for $d=3$ and large values of $\ct$ in units of the fundamental length scale $\lx^{1/4}$, 
provides a possible realization of the concept of classicalization. 
The solution describes a spherical shock front that moves in from
infinite to a minimal radius $r_{th}(0)=(4\lx \ct^2)^{1/6}$, and subsequently 
re-expands. For $\ct \gg \lx^{1/4}$ the scattering takes place at 
a length scale much larger than the fundamental one. 

There are two main difficulties in advancing this 
scenario: 
\begin{enumerate}
\item 
The numerical analysis of ref. \cite{dynamical} does not provide concrete support for the 
creation of a configuration that resembles (\ref{hz2}) starting from a purely incoming one. Shock fronts tend to 
appear during the evolution, but are of a different nature than (\ref{hz2}).
However, it must be kept in mind that the numerical approach becomes unreliable when shock waves form, so that 
it may not be able to capture the physics of the solutions discussed in this work. 
\item
It is difficult to reconcile
the apparent nonconservation of
energy and the presence of a singularity with the properties of an initially smooth, energy conserving 
configuration. A qualitative modification of the solution must occur at some point in
the evolution. We lack the analytical understanding of this process, while the numerical approach cannot
describe it reliably.
\end{enumerate}
A final possibility is that the configurations (\ref{hz2}) cannot be approached even approximately by a larger class of dynamical 
solutions with regular initial conditions. In such a case 
the process of classicalization would be very sensitive to the initial conditions.

\section{Branes with throats}\label{bt}

The DBI theory can be viewed as the effective description of a $(d+1)$-dimensional brane within 
a Minkowski bulk space with an additional spatial dimension. In this work we assume vanishing gauge fields. 
We denote the coordinates of the bulk space as 
$z^M$, $M=0,1,2,..., d+1$. We are interested in spherically symmetric configurations. For this
reason we shall mostly denote these coordinates as $t$, $r$, $\theta^1$,..., $\theta^{d-1}$, $y$ in the following.
The action can be  
identified with the area swept by the brane, which can be expressed in terms of the
induced metric as 
\be
S=-\int d^{d+1} x  \sqrt{-\det\left(\eta_{MN}\partial_\mu z^M \partial_\nu z^N \right)}.
\label{aplang} \ee
Here $x^{\mu}$, $\mu=0,1,...,d$ are the worldvolume coordinates and $z^M(x^\mu)$ the embedding functions. 
The action is invariant under arbitrary changes of the worldvolume coordinates. We can fix this gauge freedom by 
identifying $x^\mu$ with the first $d+1$ spacetime coordinates. This choice is usually characterized as 
the static gauge. The coordinate $y\equiv \phi$ becomes a field of the worldvolume theory.
The action can be written as
\be
S=-\int d^{d+1} \sqrt{1- \left(\partial_\mu\phi \right)^2}. 
\label{lagrangiandbi} \ee
This is the action resulting from the Lagrangian density (\ref{lagrangian}), 
if the fundamental scale $\lx$ is set equal to 1 and the
constant term is omitted. The theory now includes an effective cosmological constant term.
The energy density for a brane with $\phi=0$ is equal to 1. 

Our solution (\ref{hz2}) can be used in order to construct a
time-dependent generalization of the static catenoidal configuration of \cite{callan,gibbons}.   
The configuration is obtained by joining the two branches with opposite signs in eq. (\ref{hz2})
in order to create a smooth surface, which for $d>1$ looks like a throat or wormhole 
connecting two asymptotically parallel branes. The radius of the throat 
 $r_{th}(t)=\sqrt{(4 \ct^2)^{1/d}+t^2}$ is time dependent and has a minimal value that can be different
from the fundamental length scale. For $d>1$ the maximal distance between the two branes is attained for $r\to \infty$ and is time
independent. It is equal to $2D_{dt}$, with 
\be
D_{dt}=B_d \, (2\ct)^{1/d},
\label{distancet} \ee
where
\be
B_d=\frac{\sqrt{\pi}}{2d}\, \frac{\Gamma\left( \frac{d-1}{2d}\right)}{\Gamma\left( \frac{2d-1}{2d}\right)}.
\label{bd} \ee
For $d=1$ the distance between the branes increases logarithmically with $\sqrt{r^2-t^2}$.

The energy-momentum tensor can be obtained from section \ref{solut}. It is given by eq. (\ref{tmn}) by setting $\lx=1$ and 
subtracting $\eta^{\mu\nu}$ in order to account for the presence of a cosmological constant in the action (\ref{lagrangiandbi}).
The energy density and flux, as well as the
radial and angular pressures, are given by eqs. (\ref{rhodd})-(\ref{pthdd}) if we set $\lx=1$ and omit the 
terms equal to $\pm 1$ in the expressions for $\rho$, $p_r$ and $p_{\theta}$.
As a result, the energy of the total configuration is equal to $2 \Et_{th}$, with 
\be
\Et_{th}=E_{th}+\Omega_{d-1} \int_{r_{th}(t)}^\infty  r^{d-1} dr
\label{ethdiv} \ee
and $E_{th}$ given by eq. (\ref{ethroatdd}).
Clearly the energy is infinite. However, it is obvious how to renormalize it. 
We can subtract the energy of two parallel branes in the absence of a throat. The corresponding energy density of each brane 
is constant, equal to 1, while the volume results from a radial integration between 0 and $\infty$.  
The renormalized energy is $2 \Et_{th,r}$ with 
\be
\Et_{th,r}=E_{th}-\Omega_{d-1} \int_0^{r_{th}(t)}  r^{d-1} dr=0,
\label{ethren}  \ee
where we have used eq. (\ref{ethroatddd}) with $\lx=1$.

It is remarkable that 
the configuration of the two branes connected by an evolving throat has exactly
the same classical energy as the two disconnected branes. 
Even though the two configurations are not continuously connected, and probably are separated by an energy barrier, 
no additional energy is required in order to create the throat.
Our result also resolves the issue of the conservation of energy discussed in the previous section.
The increase in energy of the two connected branes is a consequence of the reduction of their worldvolume.
In other words, the expansion of the throat turns worldvolume into energy that is distributed on the branes.

It is also interesting to compare with the energy of the static throat constructed
from the two branches of eq. (\ref{stat2}) for $\lx=1$. 
The most narrow part of the throat has a radius $r_{sth}=c^{1/(d-1)}$. For $d>2$ the maximal distance between the two
branes is attained for $r\to\infty$. This distance is equal to $2D_{ds}$, with
\be
D_{ds}=A_d\, c^{1/(d-1)},
\label{dist} \ee
where
\be
A_d=\frac{\sqrt{\pi}}{2d-2}\, \frac{\Gamma\left( \frac{d-2}{2d-2}\right)}{\Gamma\left( \frac{2d-3}{2d-2}\right)}.
\label{aa} \ee
After renormalization, the total energy associated with the static throat is $2\Et_{sth,r}$, where
\be
\Et_{sth,r}=A_d \,\frac{\Omega_{d-1}}{d} c^{d/(d-1)}=A_d \,\frac{\Omega_{d-1}}{d} r^{d}_{sth}.
\label{energythr} \ee
For $d=2$ the distance between the branes 
increases logarithmically with $r$ and the configuration has infinite energy.
For $d=1$ the distance increases linearly with $r$ and the energy again diverges.
In all dimensions the static throat configuration requires more energy than the one with an evolving throat or the
two disconnected branes.

It is possible to interpret the static solution as some type of sphaleron with energy equal to the height of the 
barrier separating two unperturbed parallel branes and the same branes connected by the time-dependent throat. 
We could also identify the instanton solution that would mediate the quantum tunneling between the two 
configurations. As we mentioned in section \ref{solut}, the analytic continuation to imaginary time turns the 
time-dependent solution to a static solution with an additional spatial dimension. The action of the instanton in 
the (d+1)-dimensional theory 
is equal to the energy of a static throat configuration in a theory of (d+1)+1 dimensions. 
It can be read from eq. (\ref{energythr}), through the substitution $d\to d+1$. 

We should note that a tunneling process with the spontaneous nucleation of the throat configuration 
may not be relevant for the annihilation of branes if these 
are identified with the D-branes of string theory. As pointed out in ref. \cite{callan},
the annihilation of two D-branes can take place through string processes at a rate faster than the one 
associated with the nucleation of a throat and its subsequent growth. 

\section{A different gauge}\label{gaugee}

The singularity that appears at the location of the throat is a consequence of the ``static gauge" that we employed in
the previous section. We would like now to eliminate this singularity by employing a different gauge, in order to study
the effect on the energetics of the brane. 
We discuss this issue in the context of the (2+1)-dimensional theory, for which an exact analytical 
treatment is straightforward. The picture is very similar in 3+1 dimensions.

For $d=2$, $\lx=1$ and through the definition $\phi=\sqrt{2\ct}\,\phit$ we can rewrite eq. (\ref{hz2}) as
\be
\frac{d\phit}{du}=\pm \frac{1}{\sqrt{4u^3-4u}},
\label{eomweis} \ee
where $z=2 \ct\, u$, $t^2=2\ct\, \tit^2$, $r^2=2\ct\, \rt^2$, $u=\rt^2-\tit^2$.
The solution of this equation can be expressed in terms of the inverse Weierstrass function ${\cal P}^{-1}$ (which we take to be
positive) as 
\be
\phit(u)=\pm({\cal P}^{-1}(u,4,0)-B_2),
\label{weirz} \ee
with $B_2=\sqrt{\pi}\,\Gamma(1/4)/(4\,\Gamma(3/4)) \simeq 1.311$, as given by eq. (\ref{bd}).
We have adjusted the integration constant so that $\phit(1)=0$.

If the solution is assumed to describe a shock wave in the context of the theory (\ref{lagrangian}), the front is located  
at $z_{th}=r_{th}^2(t)-t^2=2 \ct$.
The energy density $\rho$ and radial energy flux $j_r$ are given by eqs. (\ref{rhodd}), (\ref{jjdd}) with $d=2$. 
They satisfy the continuity equation 
$\partial\rho/\partial t+(1/r)\partial(r j_r)/\partial r=0$.
At a given time $t$, the total energy of the configuration is  
\be
E_{th}=\int_{r_{th}(t)}^\infty 2\pi r \rho(t,r) dr=
\pi (2 \ct+t^2)=\pi\,r^2_{th}(t).
\label{ethroat2d} \ee
The energy is finite, but time dependent. 

In the context of the brane picture we join the two branches of eq. (\ref{weirz}) so as to
obtain two branes connected by a circular throat of radius $r_{th}(t)$. 
The maximal distance between the two branes, attained for $r\to \infty$, is $2D_{2t}$ with
\be
D_{2t}=B_2\, \sqrt{2\ct}.
\label{diste2d} \ee 
The energy density of the effective theory (\ref{lagrangiandbi}) with $d=2$ contains a cosmological constant.
As a result, the energy of the total configuration is equal to $2 \Et_{th}$, where
\be
\Et_{th}=E_{th}+2 \pi \int_{r_{th}(t)}^\infty  r\, dr
\label{ethdiv2} \ee
and $E_{th}$ given by eq. (\ref{ethroat2d}).
Through the subtraction of the energy of two parallel branes in the absence of a throat we get the
renormalized energy $2 \Et_{th,r}$ where
\be
\Et_{th,r}=E_{th}-2\pi \int_0^{r_{th}(t)}  r\, dr=0.
\label{ethren2}  \ee

An alternative way to regularize the energy is to subtract the (infinite) energy of our solution at a given time, let's say $t=0$. 
In this way we would find that the renormalized energy is $2\Et_{th,r}$, with
\be
\Et_{th,r}=\pi t^2.
\label{lal} \ee
This result reproduces the time dependence of the energy associated with the shock wave in eq. (\ref{ethroat2d}). 
The regularization is very similar to the addition of a constant term, as was done in eq. (\ref{lagrangian}) 
in order to subtract the
effective cosmological constant.
However, this regularization fails to account for the part of the worldvolume of the unperturbed branes that is bounded by the
throat. For this reason it does not remove the time dependence of the renormalized energy.

We would like to repeat the above calculation in a different gauge by choosing a different set of worldvolume coordinates. 
To this effect, we set $x^0=t$, $x^1=y$, $x^2=\theta$ and interpret $r\equiv \phi(t,y)$ as a field of the worldvolume theory.
The action can be derived starting from eq. (\ref{aplang}). It is 
\be
S=-2\pi \int dt dy \, \phi(t,y)\,\sqrt{1-\left(\partial_\mu \phi\right)^2}.
\label{lagrangiangauge} \ee
As we are interested in describing a brane configuration with cylindrical symmetry around the $y$-axis, 
we have integrated over the angular variable.

From the point of view of the embedding, the solution (\ref{weirz}) describes a moving surface and
provides a relation between the coordinates of the embedding space. This relation can be expressed equivalently by solving 
for $r$ in terms of $t$ and $\phi$ (which stands for the coordinate $y$ in the initial gauge). In this way we can produce 
another solution of the equations of motion of the worldvolume theory. By solving eq. (\ref{weirz}) for $r$ and making the
substitutions $\phi\to y$ and $r\to \phi$, we obtain
 \be
\phit(\tit,\yt)=\sqrt{\calP(\yt+B_2,4,0)+\tit^2},
\label{appsol2} \ee
where $\calP(z,4,0)$ is the Weierstrass elliptic function. 
We have used the same rescaling as before, as well as $y=\sqrt{2\ct}\, \yt$.
It is straightforward to check that this expression satisfies the equation of motion of the theory (\ref{lagrangiangauge}).
It describes a throat connecting two branes located at the points $\yt=\pm B_2$, where the Weierstrass function diverges. 
The minimal value of $\phit$ is obtained for $\yt=0$, where $\phi=\sqrt{2c}\, \phit=\sqrt{2c+t^2}$. This is the location of the
throat. 

The energy density and flux for the solution (\ref{appsol2}) are
\begin{eqnarray}
\rho&=&2\pi\sqrt{2\ct}\,\frac{4\tit^2+4 \calP(\yt+B_2,4,0)+\left[ \calP'(\yt+B_2,4,0)\right]^2}{2\sqrt{4 \calP(\yt+B_2,4,0)
+\left[ \calP'(\yt+B_2,4,0)\right]^2}}
\label{apprhob} \\
j&=&-2\pi\sqrt{2\ct}\,\frac{\tit \,\calP'(\yt+B_2,4,0)}{\sqrt{4 \calP(\yt+B_2,4,0)+\left[ \calP'(\yt+B_2,4,0)\right]^2}},
\label{appjjb} \end{eqnarray}
respectively.
They satisfy the continuity equation 
$\partial \rho/ \partial t + \partial j/ \partial y=0$.
They are finite at the location of the throat. Actually, the energy flux vanishes at this point.
On the other hand, the energy density diverges at the points $\yt=\pm B_2$, where the branes are located, while the
flux takes values $\mp2\pi t$ there. The fluid velocity $j/\rho$ vanishes at $\yt=\pm B_2$, but the presence of a
finite energy flux implies that the total energy of the configuration cannot be constant. The situation is very similar
to the one in the first gauge.
However, in the current gauge the singularity has moved from the location of the throat to the location of the
branes. 

The integration of the energy density between the points $\yt=-B_2$ and $B_2$ returns an infinite result. 
As before we need to regulate the total energy by subtracting the energy of the configuration without a throat. 
However, this is not straightforward in this gauge, as the coordinate system $(t,y,\theta)$ is singular on the unperturbed
branes. A whole brane would correspond to one value of $y$. This is to be contrasted with the situation in the first gauge,
where the coordinates $(t,r,\theta)$ naturally span the unperturbed brane.
The best we can do is to subtract the (infinite) energy of our solution at a given time, let's say $t=0$. 
The regularized energy would then be given by eq. (\ref{lal}) and have a time dependence, exactly as in the
static gauge with the same type of regularization.

We can draw two main conclusions from the analysis in this section:
\begin{enumerate}
\item
The throat is not necessarily associated with a singularity of the solution. 
An appropriate choice of coordinates (gauge choice) can make the solution regular at the location of the throat, even though it may
induce singularities at other points. We also point out that the effective action has a different form in various gauges.
For example, eq. (\ref{lagrangiangauge}) does not have the standard DBI form obtained in the static gauge.  
\item
A correct regularization of the energy of the configuration must account for the worldvolume of the branes that is eliminated 
through the throat expansion and reappers as energy distributed over the remaining part of the branes. 
This point is not apparent if eq. (\ref{lagrangian}) is used for the definition of the theory, without reference to 
the brane picture.
\end{enumerate}

\section{Conclusions}\label{concl}

The aim of this work was to understand the nature of the solution (\ref{hz2}) in the context of the DBI theory. This
theory is taken as the prototype for a class of higher derivative theories ($k$-essence,  
DGP model, Galileon, scalar-tensor models with kinetic gravity braiding etc) that have attracted considerable interest recently.
In this context the solution describes a shock wave of infinite energy density \cite{heisenberg}. 
A peculiarity of the solution is that integrating the energy density at a given time, starting from the location of the shock front, 
results in a time-dependent total energy for the configuration. On the other hand, the 
energy-momentum tensor of the theory is conserved. Reconciling these facts means that the singularity 
associated with the shock wave must be viewed as a source of energy for the region in its wake. It is 
not obvious how to introduce such a source in the formalism starting from the Lagrangian density. 

The picture can be clarified by employing the geometrical picture, in which the DBI theory describes a minimal surface in
Minkowski space with an additional spatial dimension. The solution (\ref{hz2}) then corresponds to a pair of branes connected
by a time-dependent throat. The energy of the configuration is infinite, but vanishes when renormalized with respect to the
energy of two unperturbed branes. What is viewed as the time-dependent energy of the shock-wave configuration 
corresponds to the worldvolume of the branes that is eliminated through the expansion of the throat. 
In other words, the expansion of the throat turns worldvolume into energy that is distributed on the branes.
Even within the geometrical picture, this connection is not always apparent. Using a different definition of worldvolume
coordinates (going to a different gauge) may obscure the connection of the energy with the eliminated brane worldvolume.

The DBI theory can be extended through the inclusion of other classes of higher-derivative terms in the action. 
Similarly, the brane action can be extended through the addition of terms involving the extrinsic or intrinsic curvatures.
It is remarkable that a connection can be built 
between a certain higher-derivative theory, characterized as the 
Galileon theory, and the motion of a probe brane with an extended action \cite{dbigal}. 
A variation of the solution we studied exists within these extended theories as well \cite{prep}. It has
interesting applications, such as  scattering in the context of
classicalization and dynamically evolving branes in higher-dimensional cosmology.
Another possibility is to extend the solution to the multi-throat case along the lines of \cite{hoppe,gibbons}.

The most interesting generalization of this picture would be through the inclusion of gravity. If the solution is interpreted
as a shock wave, the divergence of the energy density clearly demonstrates that gravity must become important
and possibly act as a regulator.
Within the brane picture, the backreaction of the brane on the bulk geometry is neglected in the probe approximation
that we considered. Taking it into account could result in configurations similar to the 
``bubbles of nothing" that can trigger the instability of the Kaluza-Klein vacuum \cite{witten}.
Another possible interpretation of the brane construction is as a bouncing brane-Universe that contracts down to a minimal
size and subsequently expands. The effect on the expansion rate of the energy  arising from the eliminated worldvolume
is a very interesting problem.

\section*{Acknowledgments}
We would like to thank S. Abel, A. Vikman for useful discussions
and the CERN Theory Division for hospitality during the beginning of this study.
This research has been supported in part by
the ITN network ``UNILHC'' (PITN-GA-2009-237920).
This research has been co-financed by the European Union (European Social Fund – ESF) and Greek national 
funds through the Operational Program ``Education and Lifelong Learning" of the National Strategic Reference 
Framework (NSRF) - Research Funding Program: ``THALIS. Investing in the society of knowledge through the 
European Social Fund".

\end{document}